# Flexibility of Ga-containing Type-II superlattice for long-wavelength infrared detection

M. Delmas[1]*, D. C. M. Kwan[1], M.C. Debnath[2], B. L. Liang[2], D. L. Huffaker[1,2]

[1] *School of Physics and Astronomy, Cardiff University, The Parade, Cardiff CF24 3AA, UK*

[2] *California NanoSystems Institute, University of California, Los Angeles, CA 90095, USA*

*corresponding author: now at IRnova AB, Sweden e-mail: marie.delmas@ir-nova.se

## Abstract

In this paper, the flexibility of long-wavelength Type-II InAs/GaSb superlattice (Ga-containing SL) is explored and investigated from the growth to the device performance. First, several samples with different SL period composition and thickness are grown by molecular beam epitaxy. Nearly strain-compensated SLs on GaSb exhibiting an energy band gap between 105 to 169 meV at 77K are obtained. Second, from electronic band structure calculation, material parameters are extracted and compared for the different grown SLs. Finally, two p-i-n device structures with different SL periods are grown and their electrical performance compared. Our investigation shows that an alternative SL design could potentially be used to improve the device performance of diffusion-limited devices for long-wavelength infrared detection.

# I- Introduction

The Type-II InAs/GaSb superlattice (Ga-containing SL) is a material of great interest for infrared imaging as it offers many advantages including a tuneable energy band gap in the midwave (MWIR, 3 - 5 µm) and longwave (LWIR, 8 - 12 µm) infrared spectral domain, a high absorption coefficient and low tunnelling currents. Recently, the InAs/InAsSb SL (Ga-Free SL) has emerged as an alternative to the Ga-containing SL as longer minority carrier lifetime have been reported [1][2]. Although significant results have been obtained [3][4], it is understood that the InAs/GaSb SL remains the material of choice for LWIR detection thanks to better optical and hole transport properties [5][6]. For this reason, as well as the growing interest in space applications such as Earth Observation missions [7][8], the research in recent years has mainly been focused on developing InAs/GaSb SL for LWIR detection [9][10][11][12]. A variety of engineered heterostructures, also called barrier structures, such as nBn [13], pBp [14], pBn [15], p-π-M-n [16], pBiBn [17], CBIRD [18] have been demonstrated. These devices allow for reduction of the dark-current compared to standard p-i-n photodiodes by suppressing the generation-recombination (G-R) current in the absorption region thanks to the insertion of a high band gap material [19]. For all these structures and for most of the LWIR SLs reported in the literature, the absorber layer is made of a SL composed of X monolayers (MLs) of InAs and 7 MLs of GaSb with X varying from 13 to 15 MLs, which exhibits a cut-off wavelength between 9 to 10.8 µm at a temperature of 77K. It is somehow surprising to see such a limited choice of the SL composition throughout the literature knowing that the SL offers a great flexibility in the choice of the SL period to target a specific cut-off wavelength. Indeed, the energy band gap of the SL can be tuned by varying not just the period thickness but also the average composition, that is to say it depends on the ratio R of the layer thicknesses (with R = InAs thickness / GaSb thickness per period). The material properties, i.e. electronic band structure, of the SL are ruled by the ratio R and it has already been proven that it has an influence on the electro-optical properties of midwave SL photodiodes with the same energy band gap [20][21]. The objective of this work, therefore, is to explore and investigate the period flexibility of Ga-containing SLs for the LWIR spectral range. First, the growth by molecular beam epitaxy (MBE) of various SLs with different ratio R is studied since the lattice mismatch Δa/a between the SL layer and GaSb substrate directly depends on the period composition and thickness. To compensate for the tensile strain of the InAs layer on GaSb (Δa/a ~ -0.6%), the migration-enhanced epitaxy (MEE) technique was used to grow an intentional InSb layer at the interfaces. The structural and optical properties are then evaluated by means of X-Ray Diffraction (XRD) and photoluminescence (PL) measurements. Secondly, to gain insight and understanding of the SL material, an 8-band **k·p** solver is used to calculate the dispersion curve ($E(k)$ plot) of Ga-containing SL. After briefly presenting the method, the **k·p** modeling is calibrated by comparing the calculated and measured band gap of the grown SLs. The effective mass is then extracted from the electronic band structure and discussed for different SL designs. Finally, two different p-i-n device structures are grown with an active region made of a 14 MLs InAs / 7MLs GaSb SL and a 12 MLs InAs / 4 MLs GaSb SL. Both samples show a cut-off wavelength between 10 and 11 µm at 77K. Photodiodes are fabricated, and the electrical performance compared.

## II- Growth and material characterization

1. Experiment details

All Ga-containing SL structures presented in this paper were grown on a quarter of two-inch p-type (0 0 1)-oriented GaSb substrate in a Veeco Gen 930 MBE reactor equipped with dual filament SUMO Knudsen effusion cells for gallium (Ga) and indium (In) and Mark V valved cracker effusion cells for arsenic (As) and antimony (Sb). The In and Ga growth rates were set to 0.3 and 0.5 ML/s, respectively. The InAs and GaSb layers were grown using a V/III flux ratio calibrated from RHEED oscillations of 1.2 and 2, respectively. The growth temperature was monitored by pyrometer and thermocouple and, calibrated with the (1 x 3) to (2 x 5) reconstruction transition on the GaSb substrate and buffer surface. Before the growth, the native oxide on the GaSb substrate was first thermally desorbed at 540ºC. The growth temperature was then lowered down to 490ºC for the GaSb buffer layer and to 410ºC for the SL layer.

To compensate for the tensile strain of the InAs layer on GaSb, the SL is grown with an InSb layer intentionally grown by MEE at both interfaces, namely the GaSb-on-InAs and InAs-on-GaSb interfaces. The particularity of the MEE technique is that both group III and V shutters are asynchronously opened in contrast to conventional MBE (shutters synchronously opened) [22][23]. It has been demonstrated that abrupt interfaces can be obtained by MEE [24][25] and, in the case of SLs, it has led to a smoother surface with improved optical properties compared to an InSb interface layer grown by MBE [26]. The shutter sequence used for the SL growth is therefore as follows: after the growth of the InAs layer, only the In shutter remains open. It is then closed and, only the Sb shutter is opened for 6 seconds (s) to saturate the In surface with Sb at the GaSb-on-InAs interface. The Ga shutter is then opened to grow the GaSb layer. At the InAs-on-GaSb interface, the Sb shutter remains open for an additional 6 s and, then only the In shutter is opened before growing the InAs layer. Note that the In shutter opening time is the same at both interfaces and depends on the SL period. In this work, it is used to control the thickness of the InSb interface layer. In one SL period, an InSb thickness of approximately 10% of the InAs thickness is required to compensate for the tensile strain of the latter. It is also worth mentioning that all fluxes are kept constant during the entire SL growth.

The samples presented in this paragraph consist of a 45 nm undoped GaSb buffer, followed by 100 pairs of undoped InAs/GaSb SL sandwiched between two AlSb barriers (~ 20 nm thick). Finally, an undoped GaSb capping layer is grown with a thickness of 4 MLs. The SL periods studied are made of X MLs of InAs and Y MLs of GaSb with X = 10, 12, 14 and Y = 4, 7. For the sake of clarity, we use the notation X/Y SL. The In shutter opening time is set at 1 s, 1.5 s and 2 s for a SL with an InAs layer of 10, 12 and 14 MLs, respectively.

Following the growth, the structural quality of the samples was evaluated using a Bede D1 X-ray diffractometer. To access the optical properties of the SL structures, samples were then loaded in a cryostat equipped with $CaF_2$ windows to carry out PL measurements. The samples were optically excited using a 735 nm laser diode modulated at a frequency of 20 kHz. To collect the signal, a Nicolet

iS50R Fourier transform infrared spectrometer equipped with KBr beamsplitter and MCT-A detector was used.

2. XRD and PL measurements

The XRD spectra of the ω/2θ scan around the GaSb (0 0 4) reflection of a 10/4 SL (R = 2.5), 12/4 SL (R = 3), 14/4 SL (R = 3.5) and 14/7 SL (R = 2) are presented Figure 1. The lattice mismatch Δa/a and the full-width at half maximum (FWHM) of the first-order SL satellite peak ($SL_{-1}$) extracted from the XRD spectra are summarized in Table 1. It can be seen that the 10/4 SL sample is under compressive strain on GaSb with a lattice mismatch of about Δa/a ~ 0.149%. Considering that the InSb binary is on compressive strain on GaSb (Δa/a ~ + 7.8%), this indicates that the total thickness of InSb within the period is too thick to obtain a strain-compensated SL. This result is similar to what have been obtained for midwave 7/4 SL with InSb interface grown by MEE [27]. Further adjustment of the In and Sb shutter opening times could be made to reduce the thickness of the InSb interface layers, although it has been shown in Ref. [27] that exposing the InAs layer to an Sb incident flux to form an "InSb-like" interface via Sb-for-As exchange suffices to obtain a strain-compensated SL in the case of thinner InAs layer. Nevertheless, when the InAs thickness increases the compressive strain caused by the InSb interface reduces and nearly strain-compensated SLs are obtained. In addition to having the smallest lattice mismatch (nearly ~ 0%), the 14/4 SL also has the smallest FWHM (38 arcsec) among the X/4 SLs suggesting that this sample has the best structural quality. When increasing the GaSb thickness to 7 MLs, the FWHM is increased to 58 arcsec for the 14/7 SL. However, both 14/Y SLs are nearly strain-compensated on the substrate and present the lowest FWHMs compared to the 10/4 SL and 12/4 SL.

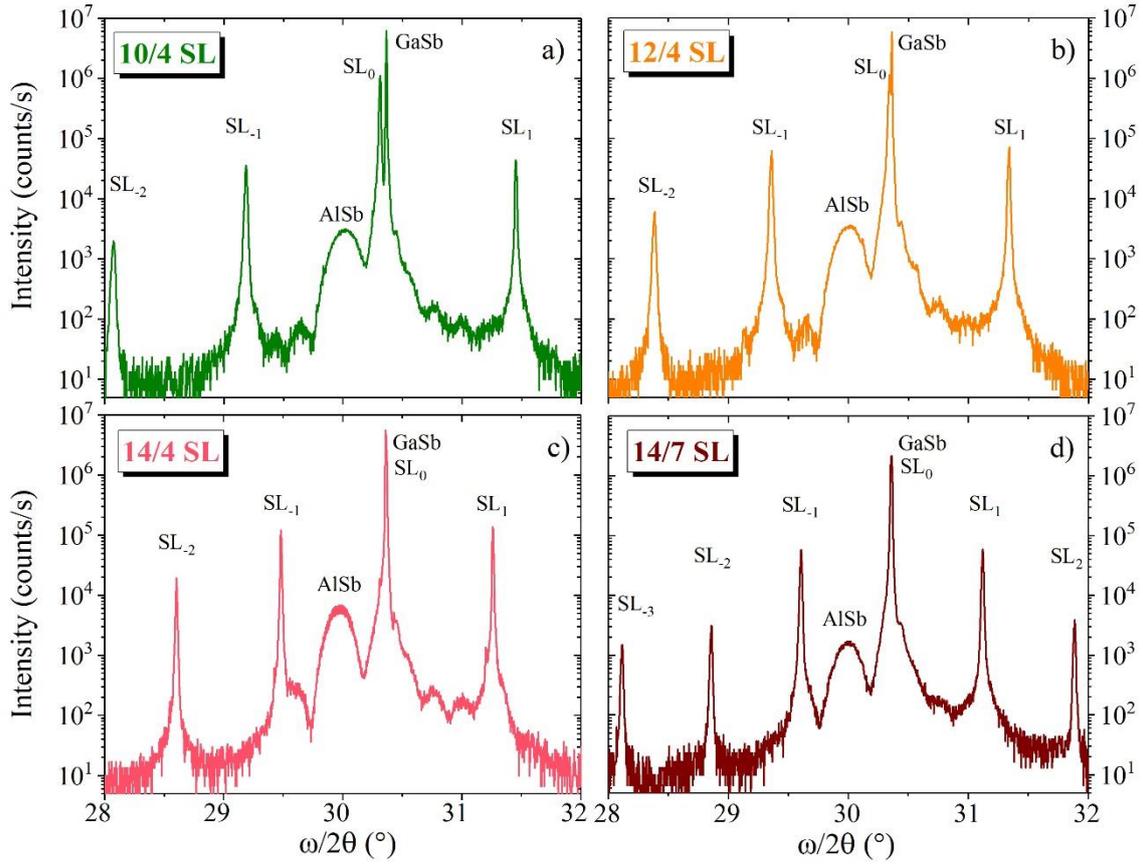

*Figure 1: XRD spectra of the sample made of a (a) 10/4 SL, (b) 12/4 SL, (c) 14/4 SL and (d) 14/7 SL.*

*Table 1: Parameters extracted from the XRD spectra and PL measurements at 77K.*

| InAs/GaSb SL [MLs] | $\Delta a/a$ [%] | FWHM $SL_{-1}$ [arcsec] | $E_g$ [eV] | PL FWHM [meV] | Max PL intensity [a.u.] |
|---|---|---|---|---|---|
| 10/4 SL | + 0.149 | 63 | 0.152 | 28 | 0.754 |
| 12/4 SL | + 0.06 | 64 | 0.124 | 28.4 | 0.556 |
| 14/4 SL | ~ 0 | 38 | 0.109 | 32.6 | 0.535 |
| 14/7 SL | ~ 0 | 58 | 0.123 | 22 | 0.424 |

The PL spectra obtained at a temperature of 77K are reported in Figure 2 (a) and (b) for the X/4 SLs and the 14/Y SLs, respectively. Note that for the sake of comparison, the PL spectrum of the 14/4 SL is plotted in both (a) and (b). Table 1 gives the energy band gap defined as the energy at the maximum PL intensity from which the thermal contribution $k_B T/2$ (with $k_B$ the Boltzmann constant and $T$ the temperature) is subtracted which corresponds to the 50% cut-off wavelength $\lambda_c$ [20], the FWHM of the PL envelope and the maximum PL intensity. We can easily observe a red shift of the band gap when the InAs thickness increases while the band gap blue shifts when the GaSb thickness increases. This results from the variation of the SL band edges depending on the period composition and thickness. This will be further discussed in section III-. In addition, we can see an increase in the FWHM of the PL envelope with the InAs thickness (X/4 SLs). One possibility is that the interface quality degrades with increasing InAs thickness. Indeed, the tensile strain induced by the InAs layer is the largest when its thickness is equal to 14 MLs, so the thickness of InSb required to compensate it is also the largest (~ 0.7 MLs at each interface). So, although a nearly strain-compensated 14/4 SL has been obtained, the material or interface quality may have been degraded as is the case in Ref. [28] for a thicker InSb interface layer. However, it appears that the PL peak is narrower for a 14/7 SL for which the InSb interface layer is relatively small compared to the total period thickness. We assume that the interface effect has a limited impact on the PL broadening in this case. Finally, the maximum PL intensity decreases with increasing InAs thickness or decreasing GaSb thickness. This behaviour is strongly correlated with the variation of the electron-hole wavefunction overlaps that have been calculated for each SL using the model described in section III- (Table 2). In addition, it is worth mentioning that additional measurements such as temperature dependent PL (from ~ 4K up to room temperature) or laser power dependent PL, for example, along with a rigorous analysis of the PL peak would provide further information about the nature of the contributing mechanisms involved in the PL emission. This is out of the scope of the present paper but will be investigated.

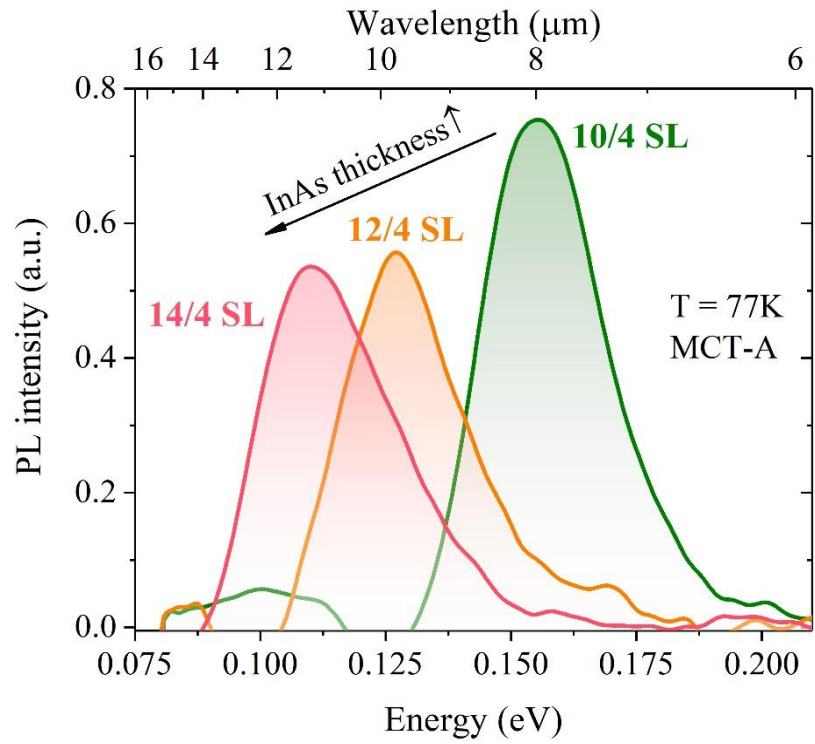

(a)

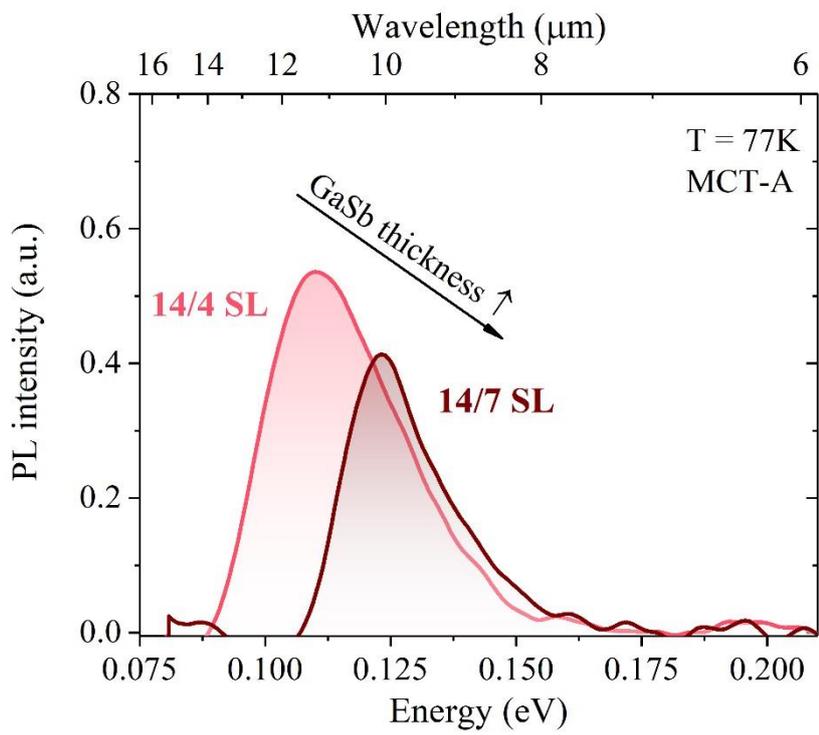

(b)

*Figure 2: Photoluminescence spectra measured at a temperature of 77K for (a) X/4 SLs and (b) 14/Y SLs.*

# III- Electronic band structure calculation

1. Method

The 8-band **k·p** envelope-function method employed for this work is available in Nextnano³ software [29] and described in detail in Ref. [30]. The interface matrix $H_{IF}$ formulated by P.C Klipstein to model the *no-atom-in-common* InAs/GaSb SL is implemented in the software framework and defined as [31]:

$$H_{IF} = \sum_i \delta(z - z_i) \begin{bmatrix} D_S & 0 & 0 & \pi_i\beta \\ 0 & D_X & \pi_i\alpha & 0 \\ 0 & \pi_i\alpha & D_X & 0 \\ \pi_i\beta & 0 & 0 & D_Z \end{bmatrix} \quad (1)$$

where $i$ is the index of the interface at the position $z_i$ and $\pi_i$ takes a value of -1 or 1 at the InAs-on-GaSb and GaSb-on-InAs interfaces. The interface parameters $\alpha$ and $\beta$ have been fixed to a value of 0.2 eV·Å [32] whereas the $D$ diagonal interface parameters ($D_S$, $D_X$, $D_Z$), which are equal to zero in the case of a common atom superlattice, are determined in order to obtain a good agreement between the calculated and measured energy band gap of the X/4 and 14/Y SLs at 77K (Figure 2). A strain-compensated SL on GaSb is assumed in our simulation by including the InSb intentional layer at both interfaces and by considering homogenous strain for the strain calculation. As previously mentioned, in one period of SL the total thickness of the InSb layer required is 10% of the InAs thickness. The material parameters for InAs, GaSb and InSb binaries used for the **k·p** band structure calculation are given in Table 1 of Ref. [33].

2. Results and Discussion

The calculated cut-off wavelength as a function of the measured cut-off wavelength at 77K is represented in Figure 3, along with the ideal prediction line. The $\pm k_B T$ deviation in the predicted $\lambda_c$ is also represented. Note that the $D$ diagonal interface parameters ($D_S$, $D_X$, $D_Z$) of Eq. (1) used are equal to (0.8, 0.3, -0.3). We observe a good agreement between the simulation and experiment for all the different SLs with an error in the $\pm k_B T$ deviation range, apart from the 10/4 SL. This can be explained by the fact that the 10/4 SL is under compressive strain on GaSb with a large lattice mismatch as discussed in section II- while in our simulation we assume a SL layer lattice matched on GaSb. In addition, a slight change in the growth rate during the MBE growth can lead to a slight change in composition and thickness of the SL period compared to the targeted layer thicknesses which we do not take into consideration in our simulation. In addition, the cut-off wavelengths calculated without taking into account the interface matrix $H_{IF}$ and considering neither the $H_{IF}$ nor the InSb layer at the interfaces are also plotted in Figure 3 for comparison. We can see that if both the $H_{IF}$ and the InSb interfaces are not considered the model cannot predict the measured cut-off wavelength and underestimates it. This result demonstrates the importance of the interface consideration for band gap calculation of Ga-containing SL.

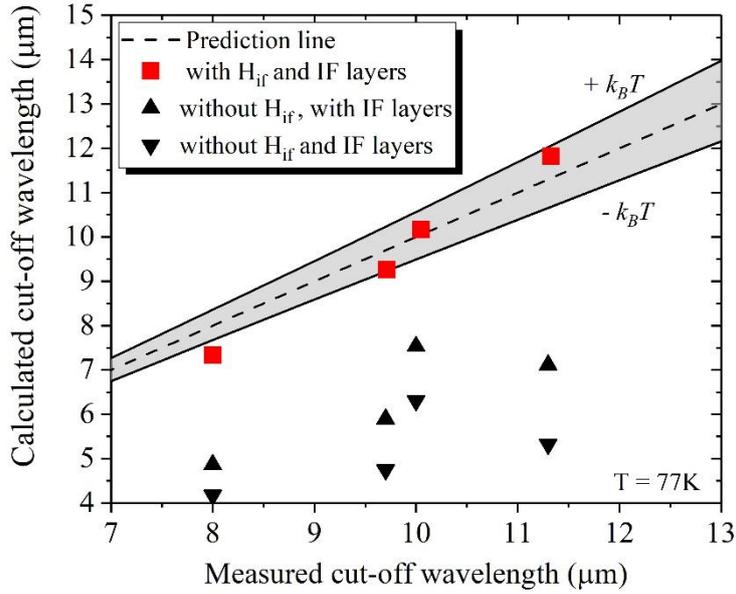

*Figure 3: Comparison between the calculated and measured $\lambda_c$ at 77K for the different SL periods (red squares) along with the ideal prediction line (dashed line). $\lambda_c$ calculated without $H_{IF}$ (triangles) and considering neither $H_{IF}$ nor the InSb layers (circles) are also plotted for comparison. The $\pm k_B T$ deviation in the predicted cut-off is represented by the solid lines and the grey area.*

Following this, the electronic band structure of the X/4 SLs and 14/Y SLs has been calculated for one in-plane direction in the Brillouin zone $k_{//}$ and in the perpendicular direction $k_\perp$ (Figure 4). We can easily see in Figure 4 that for the X/4 SLs, the bottom of the conduction band is moving down with increasing the InAs thickness. It decreases by 49 meV when the InAs thickness increases from 10 to 14 MLs while the top of the first valence band is moving up by a value of only 15 meV. For thicker GaSb, both conduction and valence bands are moving up by 51 and 35 meV, respectively. These changes impact on the energy band gap value as previously observed experimentally. It seems that the X/4 SL band structures are quite similar in contrast to the 14/Y SLs that present differences. In particular, the lower valence bands of the 14/4 SL are further removed from the top valence band (corresponding to heavy hole) compared to the 14/7 SL. Using thinner GaSb, one could therefore further minimize/suppress Auger recombination in p-type SLs.

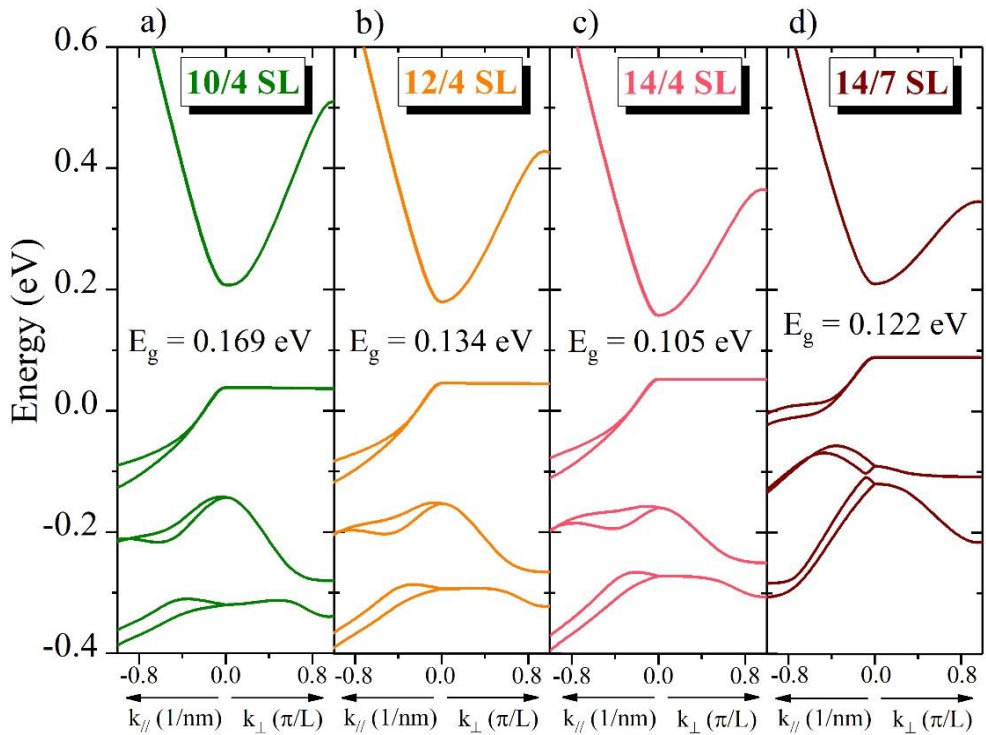

*Figure 4: Calculated electronic band structure at 77K of (a) 10/4 SL, (b) 12/4 SL, (c) 14/4 SL and (d) 14/7 SL for the in-plane direction along [100] in the Brillouin zone $k_{//}$ and in the perpendicular direction $k_\perp$ (in units of π/L with L the period thickness).*

*Table 2: Electron-hole wavefunction overlap along with the electron and hole effective masses extracted from the electronic band structure.*

| InAs/GaSb SL [ML] | Wavefunction overlap [%] | $m_e^*$ [/$m_0$] | $m_h^*$ [/$m_0$] |
|---|---|---|---|
| 10/4 SL | 63 | 0.019 | 0.176 |
| 12/4 SL | 58 | 0.018 | 0.277 |
| 14/4 SL | 54 | 0.018 | 0.303 |
| 14/7 SL | 40 | 0.022 | 0.309 |

From Figure 4, it is possible to extract the effective masses at the band edge (second derivative at the Brillouin zone center) in both directions ($m_{e,h,//}^*$ and $m_{e,h,\perp}^*$, with *e* and *h* subscripts referring to electron and heavy hole, respectively). The electron and hole effective masses calculated as $m_{e,h}^* = {m_{e,h,//}^*}^{2/3} \cdot {m_{e,h,\perp}^*}^{1/3}$ are reported in Table 2. The conduction band is similar for the X/4 SLs resulting in similar electron effective mass value, in contrast to $m_h^*$ which is strongly dependent on the InAs thickness. Indeed, the valence band width is reduced due to stronger localization of holes when increasing the InAs well thickness especially in the growth direction which results in larger effective mass in this direction. The localization of carriers is further increased, as suggested by the electron-hole wavefunction values, when increasing the GaSb layer thickness leading to larger effective mass for both electron and hole for the 14/7 SL. From Table 2, it appears that the effective masses depend mainly on the SL period composition and thickness and slightly on the energy band gap, which is not the case for bulk materials, as also illustrated in Figure 5 where the effective electron mass for different SL designs is plotted as a function of the energy band gap. The calculated value of $m_e^*$ is in the range of (0.015 to 0.040)$m_0$ for SLs with layer thicknesses ranging from 4 and 16 MLs. In comparison, the electron effective mass in the HgCdTe material is varying with the energy band gap as $0.071 \times E_g$ [34] which is smaller than in Ga-containing SLs (Figure 5). The tunnelling probability is exponentially dependent on the effective mass, the tunnelling contribution in SLs is thus lower than in HgCdTe, especially in the LWIR range where the effective mass in HgCdTe is very small compared to the one in SL. Furthermore, Figure 5 highlights the great flexibility that the InAs/GaSb SL material offers. In particular, it is possible to achieve the same energy band gap with different SL periods which have different electronic band structures and therefore, different intrinsic properties. As mentioned in the introduction, this can have an influence on the electro-optical properties of the absorber layer in a device structure. In the following section, we compare the electrical performance of two different p-i-n device structures made of a 14/7 SL and a 12/4 SL.

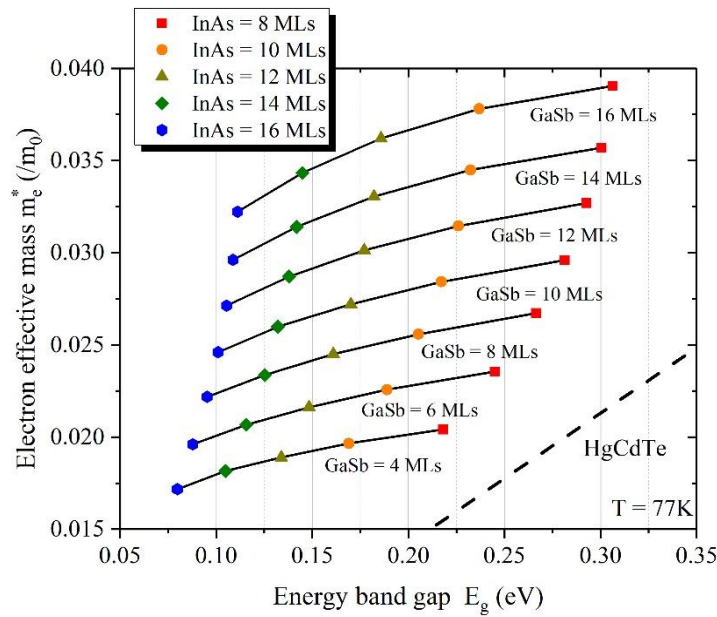

*Figure 5: Effective electron mass as a function of the energy band gap calculated for different SL designs at 77K. The horizontal solid lines are for a fixed GaSb layer thickness varying from 4 to 16 MLs. The same vertical symbols represent data for a fixed InAs layer thickness varying from 8 to 16 MLs. For comparison, the effective electron mass in HgCdTe material is also represented.*

## IV- 14/7 SL versus 12/4 SL as active region of a p-i-n photodiode

The p-i-n structures consist of a 100 nm thick p-doped GaSb buffer layer, followed by a 60 nm thick p-doped SL, a 1 µm thick non-intentionally doped (n.i.d) SL active region, a 60 nm thick n-doped SL and a 20 nm n-doped InAs cap layer. The doping concentration of the n and p contacts is $1 \times 10^{18}$ cm$^{-3}$. The InAs/GaSb SL periods studied are composed of a 14/7 SL and a 12/4 SL.

Material characterizations have first been carried out to evaluate the material quality of the two device structures. From the XRD spectra (not shown here), it appears that both samples are under a slight compressive strain on GaSb with a lattice mismatch of + 0.069 and + 0.05% for the 14/7 SL and 12/4

SL, respectively. The difference of Δa/a values with the ones reported in section II- is probably due to a slight change in the growth rate leading to a slight change in the average composition and thickness of the SL period. Nevertheless, the FWHM of SL$_{-1}$ is in the same range of about ~50 arcsec for both samples suggesting a good structural quality. PL measurements from 77K to 160K have also been performed and the energy band gap is plotted as a function of temperature in Figure 6. Its variation can be fitted using the well-known Varshni's equation which depends on the parameters α and β and the energy band gap at 0K ($E_g(0K)$). By fixing β to 270K [35], we found that $E_g(0K)$ is equal to 0.111 and 0.124 eV and, the α parameter is 0.019 and 0.094 meV/K for the 14/7 SL and 12/4 SL, respectively. In the inset of Figure 6, the PL spectra of both samples at 77K are represented. The energy band gap of the 12/4 SL (~ 0.122 eV) is close to the value previously measured while it is lower by 13 meV for the 14/7 SL (~ 0.110 eV). Again, this could originate from a slight variation in the growth rate resulting in a slight change of the SL period and therefore, of the measured cut-off wavelength.

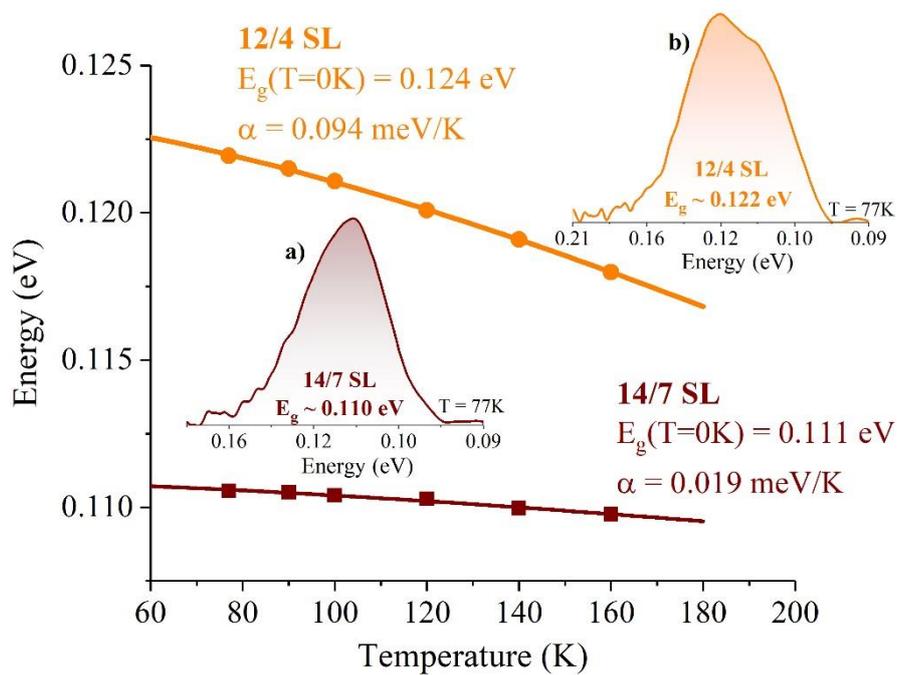

*Figure 6: Measured energy band gap as a function of temperature for the 14/7 SL (squares) and 12/4 SL (circles) device structures. The solid lines correspond to the fit with the Varshni's equation. In inset, PL spectra at 77K of (a) 14/7 SL and (b) 12/4 SL.*

From epitaxial layers, photodiodes were fabricated using standard photolithography processes. Cr/Au metal was deposited as the bottom and top ohmic contact. Mesa photodiodes with diameters ranging from 140 to 440 µm were realized by wet etching. Photoresist was then spun onto the etched surface to protect the devices from ambient air. Samples were placed in a cryogenic probe station to perform current-voltage measurements at different temperature.

The measured dark-current density at a reverse voltage of -50 mV as a function of the inverse of temperature is plotted in Figure 7 along with the variation of the diffusion current ($\propto n_i^2 \propto \exp(-E_g/k_BT)$), with $n_i$ the intrinsic carrier concentration and $E_g$ the experimental band gap from Figure 6) and G-R current ($\propto n_i \propto \exp(-E_g/2k_BT)$). At high temperature, the dark-current varies as the diffusion current for both samples. Below 100K, the G-R contribution dominates the dark-current of the 14/7 SL in contrast to the dark-current density of the 12/4 SL which cannot be fitted using the G-R variation at low temperature; we can see that it has a slower variation with temperature than the 14/7 SL. This may originate from surface leakage currents dominating the dark-current as the mesa sidewalls are not passivated or we also suspect that the 12/4 SL may be limited by tunnelling currents as it has smaller effective masses than the 14/7 SL.

In the following, we compare the dark-current density of both samples at 150K when it is limited by the diffusion current. At this temperature, the dark-current of the 12/4 SL device is lower (x 0.42) than the 14/7 SL. However, both samples show different energy band gap as illustrated in Figure 6. Therefore, for a fair comparison, we corrected the experimental dark-current by the energy band gap, i.e. we divided the dark-current by $\exp(-E_g/k_BT)$ which accounts for the diffusion component (Table 3). It appears that the 12/4 SL still shows a lower diffusion current by a factor of 0.78 than the 14/7 SL even after correction.

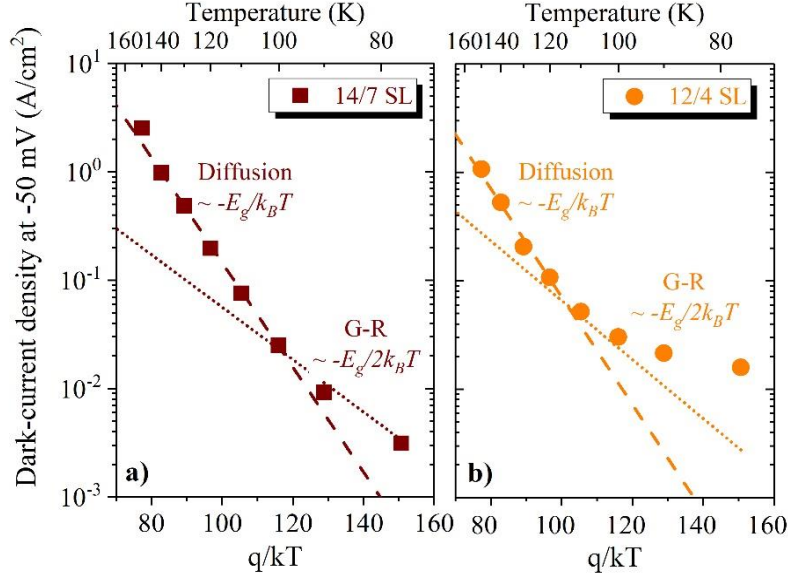

*Figure 7: Dark-current density measured at -50 mV as a function of the inverse of temperature for (a) 14/7 SL and (b) 12/4 SL p-i-n samples. The variation of the diffusion and G-R currents are represented by the dash and dot lines, respectively.*

*Table 3: Measured and corrected dark-current density at 150K for the 14/7 SL and 12/4 SL samples. The dark-current ratio is also reported.*

| $T = 150K$ | 14/7 SL [A/cm$^2$] | 12/4 SL [A/cm$^2$] | Dark-current ratio |
|---|---|---|---|
| Measured at -50 mV | 2.52 | 1.07 | 0.42 |
| Corrected by $\exp(-E_g/k_B T)$ | 12.5 x 10$^3$ | 9.8 x 10$^3$ | 0.78 |

To understand this difference, the dark-current characteristic of both samples has been simulated (Figure 8) using the Atlas framework from TCAD Silvaco software. The models and method of simulation are reported elsewhere [36]. We used the measured energy band gap and residual doping concentration of the active region as input of the simulation. It is worth mentioning that from capacitance-voltage measurements at 77K (not shown here), we extracted a residual doping concentration for the active region of around 1.4 x 10$^{15}$ cm$^{-3}$ and 1.3 x 10$^{15}$ cm$^{-3}$ for the 14/7 SL and 12/4 SL, respectively. Note that a serial resistance has also been considered in the simulation that accounts for the sheet and contact resistances, for both samples it has a value of around 1.5 Ω.cm$^2$. This only has an influence on the calculated dark-current in forward bias. The only fitting parameter of the simulation is the minority carrier lifetime $\tau$. In Figure 8, a good agreement between the simulated and experimental dark-current can be observed at 150K for positive and reverse biases. The minority lifetime extracted from the simulation is equal to 7.5 and 9.4 ns for the 14/7 SL and 12/4 SL. A relatively longer minority carrier lifetime (~ x 1.25) contributes in part to the lower diffusion current measured for the 12/4 SL device. In addition, these two SLs have different effective masses, so they have different effective density of states in the conduction $N_c$ and valence $N_v$ bands. In particular, the 12/4 SL has a lower $N_c N_v$ product by a factor of 0.63 than the 14/7 SL thanks to smaller electron and hole effective mass which also contributes to the reduction of the diffusion current. It is interesting to note that if we only account for these ratio values of minority carrier lifetime and $N_c N_v$ product, the dark-current ratio should theoretically be even lower than the measured one. This difference can be explained by other contributions not taken into account in this analysis such as surface leakage current or tunnelling current for example.

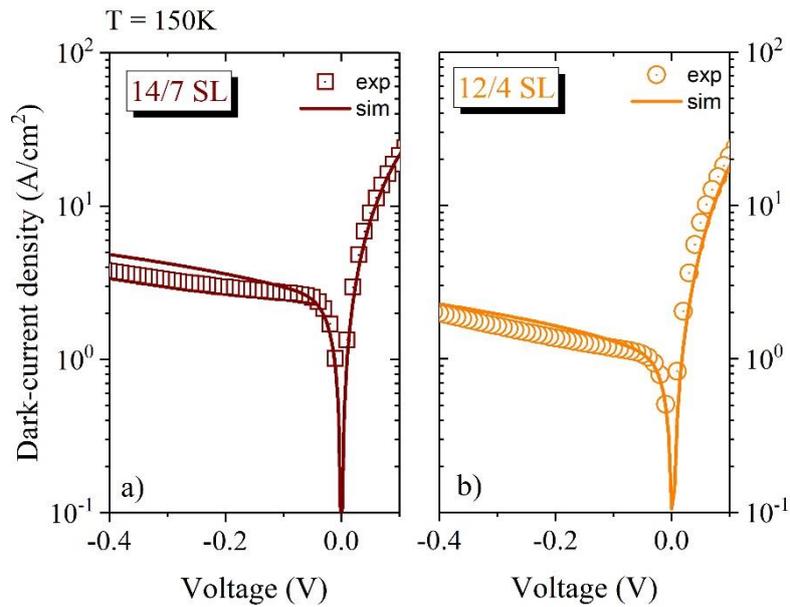

*Figure 8: Simulated (solid line) and measured (symbols) dark-current at a temperature of 150K for the (a) 14/7 SL and (b) 12/4 SL sample.*

As intermediate conclusion, due to a weaker localization of carriers, the 12/4 SL has smaller electron and hole effective masses compared to the 14/7 SL resulting in stronger tunnelling contributions in the dark-current at low temperature and a lower diffusion at high temperature. However, the former can be counteracted by using an engineered heterostructure such as pBp or nBp barrier structure which are diffusion-limited whatever the temperature. In addition, the electron-hole wavefunction overlap is enhanced in a 12/4 SL (58%) which could lead to an enhancement of the absorption coefficient, and therefore of the quantum efficiency. The present study demonstrates that the SL design can be used to improve the device performance of diffusion-limited devices. Although, we only showed a slight improvement of the diffusion current by a factor of 0.78 at 150K with a 12/4 SL, we proposed in Ref. [21] to use a 12/2 SL (~ 0.118 eV) as the absorber layer of a nBp structure to further improve the dark-current. In this case however, the growth and, in particular, the control of the interface quality (intermixing, roughness, etc.) may be more challenging.

## V- Conclusion

In conclusion, in this paper we investigated the flexibility of Ga-containing SLs. Several SL designs with a different ratio of the layer thicknesses R ranging from 2 to 3.5 that exhibit an energy band gap varying from 0.105 to 0.169 eV at 77K have been studied. First, we showed that by growing an intentional InSb interface layer using the MEE technique, nearly strain-compensated SL layers on a GaSb substrate can be obtained. The shutter opening times at the interfaces need to be precisely adjusted depending on the InAs layer thickness. The interface quality may impact on the PL broadening especially for a higher ratio R where the fraction of the interface layers is comparable to the GaSb thickness. In addition, the maximum PL intensity is strongly corelated to the electron-hole wavefunction overlap value. Second, by using an 8-band **k·p** envelope function method that can predict the measured band gap within an error of $\pm k_B T$, we calculated and compared the electronic band structure of the grown SLs. The electron effective mass has been found to be in the range of (0.015 to 0.040)$m_0$. We also demonstrated that it mainly depends on the period composition and thickness and less on the energy band gap. In particular, for a thick period with a thick GaSb layer, as is the case for a 14/7 SL, the electron and hole effective masses have been found to be the largest due to a stronger carrier localization. Finally, two different p-i-n device structures with the active region made of a 14/7 SL and 12/4 SL have been grown and photodiodes fabricated. By comparing their electrical performance, we showed that at low temperature the 12/4 SL seems to be limited by surface leakage currents or by tunnelling contributions. However, at high temperature it has a lower diffusion current compared to a 14/7 SL thanks to smaller electron and hole effective masses reducing the $N_c N_v$ product and to a relatively longer minority carrier lifetime. Furthermore, a thinner SL period leads to an enhancement of the electron-hole wavefunction overlap which could result in an enhancement of the optical properties. These results demonstrate that alternative SL designs can potentially be used to improve the device performance of diffusion-limited devices.

## Acknowledgement

This project has received funding from the European Union's Horizon 2020 research and innovation programme under the Marie Sklodowska-Curie grant agreement No 743521 (Project acronym: ASISA). The authors would also like to acknowledge the financial support provided by Sêr Cymru National Research Network in Advanced Engineering and Materials and by the EPSRC for the Future Compound Semiconductor Manufacturing Hub, reference number EP/P006973/1.

## References

[1] B.V. Olson, E.A. Shaner, J.K. Kim, J.F. Klem, S.D. Hawkins, L.M. Murray, J.P. Prineas, M.E. Flatté, T.F. Boggess, "Time-resolved optical measurements of minority carrier recombination in a mid-wave infrared InAsSb alloy and InAs/InAsSb superlattice", Appl. Phys. Lett. **101**, 092109 (2012).
[2] E.H. Steenbergen, B.C. Connelly, G.D. Metcalfe, H. Shen, M. Wraback, D. Lubyshev, Y. Qiu, J.M. Fastenau, A.W.K. Liu, S. Elhamri, O.O. Cellek, Y.-H. Zhang, "Significantly improved minority carrier

lifetime observed in a long-wavelength infrared III-V type-II superlattice comprised of InAs/InAsSb", Appl. Phys. Lett. **99**, 251110 (2011).

[3] A. Haddadi, A. Dehzangi, S. Adhikary, R. Chevallier, M. Razeghi, "Background–limited long wavelength infrared InAs/InAs$_{1-x}$Sb$_x$ type-II superlattice-based photodetectors operating at 110 K", APL Materials, **5**(3), 035502 (2017).

[4] D. Z. Ting, A. Soibel, A. Khoshakhlagh, S. B. Rafol, S. A. Keo, L. Höglund, A. M. Fisher, E. M. Luong, S. D. Gunapala, "Mid-wavelength high operating temperature barrier infrared detector and focal plane array", Appl. Phys. Lett. **113**, 021101 (2018).

[5] P.C. Klipstein, Y. Livneh, A. Glozman, S. Grossman, O. Klin, N. Snapi, E. Weiss, "Modeling InAs/GaSb and InAs/InAsSb Superlattice Infrared Detectors". J. Electron. Mater **43**, 2984 (2014).

[6] D. Z. Ting, A. Soibel, S. D. Gunapala, "Type-II superlattice hole effective masses", Inf. Phys. Technol. **84**, 102–106 (2017).

[7] L. Duvet, M. Bavdaz, P.E. Crouzet, N. Nelms, Y. R. Nowicki-Bringuier, B. Shortt, P. Verhoeve, "European Space Agency detector development for space science: present and future activities," Proc. SPIE 9154, 915403 (2014);

[8] L. Höglund, J. B. Rodriguez, S. Naureen, R. Ivanov, C. Asplund, R. Marcks von Würtemberg, R. Rossignol, P. Christol, A. Rouvié, J. Brocal, O. Saint-Pé, E. Costard, "Very long wavelength type-II InAs/GaSb superlattice infrared detectors," Proc. SPIE 10624, 106240I (2018);

[9] L. Höglund, J. B. Rodriguez, R. Marcks von Würtemberg, S. Naureen, R. Ivanov, C. Asplund, R. Alchaar, P. Christol, A. Rouvié, J. Brocal, O. Saint-Pé, E. Costard, "Influence of shallow versus deep etching on dark current and quantum efficiency in InAs/GaSb superlattice photodetectors and focal plane arrays for long wavelength infrared detection", Inf. Phys. Technol. **95**, 158–163 (2018).

[10] P. C. Klipstein, Y. Benny, S. Gliksman, A. Glozman, E. Hojman, O. Klin, L. Langof, I. Lukomsky, I. Marderfeld, M. Nitzani, N. Snapi, E. Weiss, "Minority carrier lifetime and diffusion length in type II superlattice barrier devices", Inf. Phys. Technol. **96**, 155–162 (2019).

[11] A. Soibel, D. Z. Ting, S. B. Rafol, A. Khoshakhlagh, A. Fisher, S. A. Keo, S. D. Gunapala, "Performance and radiation tolerance of InAs/GaSb LWIR detectors based on CBIRD design", Proc. SPIE 9974, 99740L (2016).

[12] R. Chevallier, A. Haddadi, M. Razeghi, "Toward realization of small-size dual-band long-wavelength infrared photodetectors based on InAs/GaSb/AlSb type-II superlattices", Solid-State Electron. **136**, 51-54 (2017).

[13] A. Khoshakhlagh, S. Myers, H. Kim, E. Plis, N. Gautam, S. J. Lee, S. K. Noh, L. R. Dawson, S. Krishna, "Long-wave InAs/GaSb superlattice detectors based on nBn and pin designs", IEEE J. Quantum Electron. **46**, 959–964 (2010).

[14] P.C. Klipstein, E. Avnon, Y. Benny, R. Fraenkel, A. Glozman, S. Grossman, O. Klin, L. Langoff, Y. Livneh, I. Lukomsky, M. Nitzani, L. Shkedy, I. Shtrichman, N. Snapi, A. Tuito, E. Weiss, "InAs/GaSb Type II superlattice barrier devices with a low dark current and high quantum efficiency", Proc. SPIE 9070, 90700U-1 (2014).

[15] A. D. Hood, A. J. Evans, A. Lkhlassi, D. L. Lee and W. E. Tennant, "LWIR Strained-Layer Superlattice Materials and Devices at Teledyne Imaging Sensors", J. Electron. Mater. **39**, 1001–1006 (2010).

[16] B. -M. Nguyen, D. Hoffmann, P. -Y. Delaunay, E. K. Huang, M. Razeghi, "Very high performance LWIR and VLWIR Type-II InAs/GaSb superlattice photodiodes with M-structure barrier", Proc. SPIE 7082, 708205 (2008).

[17] M. Gautam, H. S. Kim, M. N. Kutty, E. Plis, L. R. Dawson, S. Krishna, "Performance improvement of longwave infrared photodetector based on type-II InAs/GaSb superlattices using unipolar current blocking layers", Appl. Phys. Lett. **96**, 231107 (2010).

[18] D. Z. Y. Ting, C. J. Hill, A. Soibel, S. A. Keo, J.M. Mumolo, J. Nguyen, S. D. Gunapala, "A high-performance long wavelength superlattice complementary barrier infrared detector", Appl. Phys. Lett. **95**, 023508 (2009).

[19] S. Maimon and G. W. Wicks, "nBn detector, an infrared detector with reduced dark current and higher operating temperature", Appl. Phys. Lett. **89**, 151109 (2006).

[20] R. Taalat, J. -B. Rodriguez, M. Delmas, P. Christol, "Influence of the period thickness and composition on the electro-optical properties of type-II InAs/GaSb midwave infrared superlattice photodetectors", J. Phys. D: Appl. Phys. **47**, 015101 (2014).

[21] E. Giard, I. Ribet-Mohamed, J. Jaeck, T. Viale, R. Haïdar, R. Taalat, M. Delmas, J. -B. Rodriguez, E. Steveler, N. Bardou, F. Boulard, P.Christol, "Quantum efficiency investigations of type-II InAs/GaSb midwave infrared superlattice photodetectors", J. Appl. Phys. **116**, 1–7 (2014).

[22] B. R. Bennett, B. V. Shanabrook, R. J. Wagner, J. L. Davis, J. R. Waterman, "Control of interface stoichiometry in InAs / GaSb superlattices grown by molecular beam epitaxy", Appl. Phys. Lett. **63**, 949 (1993).

[23] H. J. Haugan, G. J. Brown, L. Grazulis, "Effect of interfacial formation on the properties of very long wavelength infrared InAs/GaSb superlattices", J. Vac. Sci. Technol. B **29**, 03C101 (2011).

[24] M. Seta, H. Asahi, S. G. Kim, K. Asami, S. Gonda, "Gas source molecular beam epitaxy / migration enhanced epitaxy growth of InAs/AlSb superlattices", J. Appl. Phys. **74**, 5033 (1993).

[25] M. Inoue, M. Yano, H. Furuse, N. Nasu and Y. Iwai, "Optical analysis of InAs heterostructures grown by migration-enhanced epitaxy", Semicond. Sci. Technol. **8**, S121-124 (1993).

[26] Z. Xu, J. Chen, Y. Zhou, Q. Xu, C. Jin and H. Li, "Interface design and properties in InAs/GaSb type-II superlattices grown by molecular beam epitaxy", Proc. SPIE 8907, 89073J (2013);

[27] M. Delmas, M. C. Debnath, B. L. Liang, D. L. Huffaker, "Material and device characterization of Type-II InAs/GaSb superlattice infrared detectors", Inf. Phys. Technol. **97**, 286-290 (2018).

[28] A. Khoshakhlagh, E. Plis, S. Myers, Y. D. Sharma, L. R. Dawson, S. Krishna, "Optimization of InAs/GaSb type-II superlattice interfaces for long-wave (~8 μm) infrared detection", J. Cryst. Growth, **311**, 1901–1904 (2009).

[29] https://www.nextnano.de/nextnano3/

[30] S. Birner, "Modeling of semiconductor nanostructures and semiconductor-electrolyte interfaces", Selected Topics of Semiconductor Physics and Technology, Vol. 135, Verein zur Förderung des Walter Schottky Instituts de Technischen Universität München e.V., München, 239 pp. (2011). ISBN 978-3-941650-35-0.

[31] P. C. Klipstein, "Operator ordering and interface-band mixing in the Kane-like Hamiltonian of lattice-matched semiconductor superlattices with abrupt interfaces", Phys. Rev. B **8**, 235314 (2010).

[32] Y. Livneh, P. C. Klipstein, O. Klin, N. Snapi, S. Grossman, A. Glozman, E. Weiss, "k·p model for the energy dispersions and absorption spectra of InAs/GaSb type-II superlattices", Phys. Rev. B **86**, 235311 (2012).

[33] M. Delmas, B. L. Liang, D. L. Huffaker, "A comprehensive set of simulation tools to model and design high-performance Type-II InAs/GaSb superlattice infrared detectors," Proc. SPIE 10926, 109260G (2019);

[34] A. Rogalski, "HgCdTe infrared detector material: history, status and outlook", Rep. Prog. Phys. **68**, 2267-2336 (2005).

[35] B. Klein, E. Plis, M. N. Kutty, N. Gautam, A. Albrecht, S. Myers, S. Krishna, "Varshni parameters for InAs/GaSb strained layer superlattice infrared photodetectors", J. Phys. D: Appl. Phys. **44**, 075102 (2011).

[36] M. Delmas, J.-B. Rodriguez, P. Christol, "Electrical modeling of InAs/GaSb superlattice mid-wavelength infrared pin photodiode to analyze experimental dark current characteristics", J. Appl. Phys. **116**, 113101 (2014).

# Table Captions

Table 1: Parameters extracted from the XRD spectra and PL measurements at 77K.

Table 2: Electron-hole wavefunction overlap along with the electron and hole effective masses extracted from the electronic band structure.

Table 3: Measured and corrected dark-current density at 150K for the 14/7 SL and 12/4 SL samples. The dark-current ratio is also reported.

# Figure Captions

Figure 1: XRD spectra of the sample made of a (a) 10/4 SL, (b) 12/4 SL, (c) 14/4 SL and (d) 14/7 SL.

Figure 2: Photoluminescence spectra measured at a temperature of 77K for (a) X/4 SLs and (b) 14/Y SLs.

Figure 3: Comparison between the calculated and measured $\lambda_c$ at 77K for the different SL periods (red squares) along with the ideal prediction line (dashed line). $\lambda_c$ calculated without $H_{IF}$ (triangles) and considering neither $H_{IF}$ nor the InSb layers (circles) are also plotted for comparison. The $\pm k_B T$ deviation in the predicted cut-off is represented by the solid lines and the grey area.

Figure 4: Calculated electronic band structure at 77K of (a) 10/4 SL, (b) 12/4 SL, (c) 14/4 SL and (d) 14/7 SL for the in-plane direction along [100] in the Brillouin zone $k_{//}$ and in the perpendicular direction $k_\perp$ (in units of π/L with L the period thickness).

Figure 5: Effective electron mass as a function of the energy band gap calculated for different SL designs at 77K. The horizontal solid lines are for a fixed GaSb layer thickness varying from 4 to 16 MLs. The same vertical symbols represent data for a fixed InAs layer thickness varying from 8 to 16 MLs. For comparison, the effective electron mass in HgCdTe material is also represented.

Figure 6: Measured energy band gap as a function of temperature for the 14/7 SL (squares) and 12/4 SL (circles) device structures. The solid lines correspond to the fit with the Varshni's equation. In inset, PL spectra at 77K of (a) 14/7 SL and (b) 12/4 SL.

Figure 7: Dark-current density measured at -50 mV as a function of the inverse of temperature for (a) 14/7 SL and (b) 12/4 SL p-i-n samples. The variation of the diffusion and G-R currents are represented by the dash and dot lines, respectively.

Figure 8: Simulated (solid line) and measured (symbols) dark-current at a temperature of 150K for the (a) 14/7 SL and (b) 12/4 SL sample.